%% file: cgo2026.tex
\documentclass[conference]{IEEEtran}
\IEEEoverridecommandlockouts
\input{preamble}
\def\BibTeX{{\rm B\kern-.05em{\sc i\kern-.025em b}\kern-.08em
    T\kern-.1667em\lower.7ex\hbox{E}\kern-.125emX}}
\newcommand{\NAME}{\xspace{\tt VecTrans}\xspace}
\newcommand{\warn}[1]{\textcolor{Red2}{}}
\newcommand{\xwz}[2]{\textcolor{black}{#1}}
\newcommand{\cl}[2]{\textcolor{black}{#1}}
\newcommand{\tyl}[2]{\textcolor{black}{#1}}

\newcommand{\wuk}[2]{\textcolor{black}{#1}}

\begin{document}

\title{\NAME: Enhancing Compiler Auto-\underline{Vec}torization through LLM-Assisted Code \underline{Trans}formations}

\author{\IEEEauthorblockN{1\textsuperscript{st} Zhongchun Zheng}
\IEEEauthorblockA{
\textit{Huawei and Sun Yat-sen University}\\
Shenzhen/Guangzhou, China \\
zhengzhongchun@huawei.com\\
zhengzhch3@mail2.sysu.edu.cn}
\and
\IEEEauthorblockN{2\textsuperscript{nd} Kan Wu}
\IEEEauthorblockA{
\textit{Sun Yat-sen University}\\
Guangzhou, China \\
wukan3@mail2.sysu.edu.cn}
\and
\IEEEauthorblockN{3\textsuperscript{rd} Long Cheng}
\IEEEauthorblockA{
\textit{Huawei}\\
Beijing, China \\
chenglong86@huawei.com}
\and
\IEEEauthorblockN{4\textsuperscript{th} Lu Li}
\IEEEauthorblockA{
\textit{Huawei HongKong}\\
HongKong, China \\
li.lu2@huawei.com}
\and
\IEEEauthorblockN{5\textsuperscript{th} Rodrigo C. O. Rocha}
\IEEEauthorblockA{
\textit{Huawei UK}\\
Cambridge, United Kingdom \\
Rodrigo.Rocha@huawei.com}
\and
\IEEEauthorblockN{6\textsuperscript{th} Tianyi Liu}
\IEEEauthorblockA{
\textit{Huawei UK}\\
Cambridge, United Kingdom \\
tianyi.liu3@h-partners.com}
\and
\IEEEauthorblockN{7\textsuperscript{th} Wei Wei}
\IEEEauthorblockA{
\textit{Huawei}\\
Hangzhou,China  \\
weiwei64@huawei.com}
\and
\IEEEauthorblockN{8\textsuperscript{th} Jianjiang Zeng}
\IEEEauthorblockA{
\textit{Huawei}\\
Shanghai,China  \\
jianjiang.ceng@huawei.com}
\and
\IEEEauthorblockN{9\textsuperscript{th} Xianwei Zhang}
\IEEEauthorblockA{
\textit{Sun Yat-sen University}\\
Guangzhou,China  \\
zhangxw79@mail.sysu.edu.cn}
\and
\IEEEauthorblockN{10\textsuperscript{th} Yaoqing Gao}
\IEEEauthorblockA{
\textit{Huawei CA}\\
Toronto,Canada  \\
yaoqing.gao@huawei.com}
}
\maketitle
\thispagestyle{plain}
\begin{abstract}
\wuk{Auto-vectorization is a fundamental optimization for modern compilers to exploit SIMD parallelism. However, state-of-the-art approaches still struggle to handle intricate code patterns, often requiring manual hints or domain-specific expertise.}{}
Large language models (LLMs), with their ability to capture intricate patterns, provide a promising solution,  \xwz{yet their effective application in compiler optimizations remains an open challenge due to issues such as hallucinations and a lack of domain-specific reasoning}{but how to best exploit LLM for compilers and overcome LLM's hallucination is still an open question}.
\wuk{In this paper, we}{This paper} present \NAME, a \xwz{novel}{} framework that \xwz{leverages}{harnesses the power of} LLMs to enhance compiler-based code vectorization. \NAME first employs compiler analysis to identify potentially vectorizable code regions. It then \xwz{utilizes}{leverages} an LLM to refactor \xwz{these regions}{the code region} into patterns \xwz{that are more amenable}{easier} to \xwz{}{recognize and realize by}the compiler's auto-vectorization. \xwz{}{pass.}To ensure semantic correctness, \NAME \xwz{further integrates}{incorporates} a hybrid validation mechanism at the \xwz{}{compiler’s}intermediate representation \xwz{(IR) level}{}.
\xwz{With the above efforts, VecTrans combines the adaptability of LLMs with the precision of compiler vectorization, thereby effectively opening up the vectorization opportunities.}{}
Experimental \xwz{results show that}{evaluation shows,} \xwz{among}{of} all\wuk{}{50} TSVC functions unvectorizable by \wuk{GCC, ICC, Clang, and BiSheng Compiler}{Clang, GCC, and BiShengCompiler}, \NAME \xwz{achieves}{achieving} \xwz{an geomean speedup of 1.77x}{1.77× geomean speedup} and successfully vectorizes 24 \wuk{of 51 test cases}{cases (46\%)}. \wuk{This marks a significant advancement over state-of-the-art approaches while maintaining a cost efficiency of \$0.012 per function optimization for LLM API usage}{}.

\end{abstract}


\section{Introduction}

Vectorization is \xwz{a critical}{an important}\wuk{}{ compiler} optimization that transforms scalar operations into vectorized forms to exploit the parallel processing capabilities of SIMD (Single Instruction, Multiple Data) architectures.
Vectorization dramatically improves \wuk{data}{computational} throughput\wuk{}{ by simultaneously processing multiple data elements as a single vector operation}, making it indispensable for performance-critical domains such as scientific computing, machine learning\xwz{, and}{,} multimedia processing\xwz{}{, etc}.
\wuk{Consequently, automatic vectorization now represents a standard feature in all mainstream compilers.}{}
\xwz{}{However, }Despite decades of \xwz{research and engineering investment}{research}, automatic vectorization still faces \xwz{serious challenges}{some problems}, \xwz{including complex data dependencies and diverse code patterns}{such as the complexity of dependencies and the diversity of code patterns}~\cite{rocha20valu,chen22}.

Currently, several strategies are \tyl{available}{employed} to address \wuk{code vectorization, especially}{} non-vectorized loops, including \wuk{manual source code modifications}{manually modifying the source code}, \wuk{intrinsic usage}{using intrinsic instructions}, and \wuk{targeted vectorization optimizations}{applying targeted vectorization optimizations}. \wuk{These methods, while functional, are individually constrained by specific drawbacks}{While these methods can be effective, they each come with their own set of drawbacks}. Manual source code modifications (e.g., rewriting loops into SIMD-friendly structures) preserves readability through standard C/C++ implementations, enabling cross-platform compatibility via compiler portability.
However, manually optimizing programs is not only labourious, but also error-prone.
Intrinsics, e.g., Intel’s {\tt \_mm512\_load\_ps}, provide precise hardware control, but sacrifice portability by binding code to specific instruction sets (e.g., AVX-512 intrinsics \wuk{are incompatible with}{fail on} ARM platforms) and obfuscating algorithmic intent. Targeted optimizations automate some vectorization tasks but struggle with non-affine loops or dynamic control flow.

Even compiler-driven approaches \xwz{confront}{face} fundamental tensions: while vanilla C/C++ code retains platform independence, reliance on compiler \wuk{(e.g., GCC, ICC, Clang\cite{lattner2004llvm})}{} auto-vectorization \wuk{may}{} \xwz{lead to}{yields} inconsistent results. The \xwz{core challenge}{root issue} lies in conflicting priorities: maintaining portable source code \xwz{requires}{necessitates} abstracting hardware details, \xwz{whereas}{yet} maximizing SIMD utilization \xwz{demands}{requires} deep architectural awareness (e.g., register widths, masking support). This dichotomy forces developers into a \wuk{}{lose-lose}choice: either accept suboptimal performance with portable scalar code or lock into vendor-specific intrinsics for marginal gains. Moreover, the \xwz{inherent limitations of}{} \wuk{the}{} heuristic-driven \xwz{nature of modern compiler}{} vectorization \xwz{often fail to fully exploit the available SIMD parallelism, leaving significant performance potential untapped}{in modern compilers often prevent the full exploitation of latent SIMD parallelism}.


These limitations collectively \xwz{highlight}{underscore} the need for a \wuk{disruptive}{} paradigm that transcends the manual-vs-automatic divide—one that preserves code portability while achieving \xwz{optimizations close to hand-tuned intrinsics}{near-intrinsic levels of optimization}. \wuk{}{Our work addresses this \xwz{challenge}{gap} by leveraging LLMs to infer architectural constraints from code semantics, enabling automated vectorization that adapts to both program logic and hardware capabilities without \xwz{compromising}{sacrificing} readability or portability.}
\wuk{LLMs, p}{P}retrained on vast code corpora, \wuk{}{LLMs} exhibit emergent capabilities in understanding complex program semantics and generating context-aware transformations. Unlike rule-based compilers, LLMs can reason \xwz{out}{about} implicit parallelism opportunities in loops with irregular structures, such as those containing conditional branches or indirect memory accesses. \wuk{}{Our approach \xwz{employs}{utilizes} a multi-\cl{feedback}{task} architecture that \xwz{simultaneously}{jointly learns to} analyzes data dependencies, predicts vectorization viability, and generates optimized loop variants that can maximize the \xwz{potential of LLM-driven code transformation}{power of LLMs}.}

However,\wuk{}{naively} applying LLMs to \cl{code optimization}{vectorization} \xwz{presents}{introduces} new challenges. Firstly, \cl{LLM is a category of probabilistic models proficient in solving some combinatorial explosion problems. Nevertheless, it suffers from serious hallucinations, especially in tasks related to strong reasoning like mathematical or code generation problems. \tyl{This unresolved question makes}{These unresolved questions make } its capability unreliable~\cite{zhou2024larger, mirzadeh2024gsm} \wuk{when}{if} utilized in \wuk{previous targeted}{our studied} auto-vectorization\wuk{}{optimization}. To make LLM's ability more robust and deterministic, a\wuk{}{solid} framework \wuk{named}{usually called} AI agent is essential to enhance the interaction of LLM with other robust and precise tools like traditional compilers. Secondly, \tyl{generating}{how to generate} both syntactically and semantically correct code is usually a fundamental requirement for auto-vectorization in traditional compilers, but \tyl{this is not guaranteed in}{the situation is not suited to} naive LLMs. }{ Secondly, the \xwz{vast}{} combinatorial space of possible loop transformations makes \xwz{}{an} exhaustive search computationally \xwz{infeasible}{intractable}.}

To \xwz{overcome these challenges}{address these issues}, we designed a verification-aware \cl{multi-feedback}{} iterative refinement \cl{}{(VAIR)} \cl{framework, \tyl{which can be considered as an AI agent}{also able to viewed as an AI Agent}, for auto-vectorization optimization}{}. \cl{The iterative refinement method is proposed by Wilkinson\cite{wilkinson1965rounding} in linear system solving problems. \tyl{It}{Simultaneously, it} is firstly introduced in LLM by Madaan\cite{madaan2023self} to remarkably enhance the capability of naive LLMs using self-refine mechanism.}{} 
Correspondingly, the framework is divided into three phases: (1) \cl{generate the candidate transformed source code using LLM}{ \xwz{first generates}{proposes}}, with the requirement that the generated code is more amenable to compiler auto-vectorization; (2) verify semantic equivalence between the original and transformed source codes utilizing unit test and, successively, \tyl{conduct the}{} formal verification with a symbolic execution engine; (3) invalid candidates trigger model retraining with counterexamples, progressively improving vectorization accuracy. This feedback loop enables the system to surpass the coverage limitations of static analysis while maintaining correctness guarantees, \xwz{marking}{which is} a critical advancement over existing \xwz{compiler-based approaches}{compilers}. Additionally, the objective of this task is to generate code that is more conducive to compiler auto-vectorization by performing source-level transformations, while ensuring the optimized code maintains readability, portability, and verifiability.


\xwz{In summary,}{} the paper makes the following contributions:
\begin{itemize}
    \item \xwz{\textbf{LLM-compiler collaboration paradigm}: We explore a new paradigm that combines the inference capabilities of LLMs with the optimization strengths of compilers. This approach enables more accurate and higher-performing vectorization.}{\textbf{Explores a new paradigm of cooperation between LLMs and compilers.} Explore higher optimization and more accurate vectorization methods based on the inference capability of large models and the optimization capability of compilers.}
    \item \xwz{\textbf{LLM-guided source code transformations for vectorization}: We introduce a novel approach that leverages LLMs to enhance vectorization opportunities through source code transformations. Our proposed method ensures cross-platform versatility while generating more human-readable code.}{\textbf{A novel approach to uncovering more vectorization} through LLM-based source code transformations. Ensure cross-platform versatility and generate code that is more human-readable.}
    \item \wuk{\textbf{Effective results and cross-platform verification}: The evaluations show that VecTrans effectively extends the vectorization cases, and further achieves significant performance improvements. Meanwhile, correctness across platforms is ensured by performing formal verification at the IR level, eliminating the need for platform-specific validation.}{}
\end{itemize}

\begin{figure}[htbp]
\centering
\begin{subfigure}[t]{0.5\textwidth}
\begin{lstlisting}[   % 进行参数设置
 language=C, % 设置语言
 basicstyle=\ttfamily, % 设置字体族
 breaklines=true, % 自动换行
 keywordstyle=\bfseries\color{DodgerBlue3}, % 设置关键字为粗体，颜色为 DodgerBlue3,
 columns=flexible
]  
void s1113(...) {
    for (int nl = 0; nl < 2 * iters; nl++)
        for (int i = 0; i < LEN_1D; i++)
            a[i] = a[LEN_1D/2] + b[i];
}  
\end{lstlisting}
\caption{Original non-vectorizable code.}
\end{subfigure}
\begin{subfigure}[t]{0.5\textwidth}
\begin{lstlisting}[   % 进行参数设置
 language=C, % 设置语言
 basicstyle=\ttfamily, % 设置字体族
 breaklines=true, % 自动换行
 keywordstyle=\bfseries\color{DodgerBlue3}, % 设置关键字为粗体，颜色为 DodgerBlue3,
 columns=flexible
]  
void s1113_opt(...) {
    for (int nl = 0; nl < 2 * iters; nl++){
        int mid = LEN_1D / 2;
        float temp = a[mid];
        for (int i = 0; i < mid; i++)
            a[i] = temp + b[i];
        a[mid] = temp + b[mid];
        temp = a[mid];
        for (int i = mid+1; i < LEN_1D; i++)
            a[i] = temp + b[i];
    }
}
\end{lstlisting}
\caption{Transformed vectorizable code generated by \NAME.}
\end{subfigure}

\caption{Example from \texttt{TSVC\_2}. The original loop cannot be vectorized due to a cross-iteration dependency on \texttt{a[LEN\_1D/2]}. \NAME transforms the loop by splitting it at the dependent index and introducing temporary variables, enabling safe vectorization.}
\label{code:s1113}
\end{figure}

\section{Background and Motivation}

\subsection{Background}

\wuk{The theoretical foundation of SIMD and vectorization dates back to Flynn’s taxonomy of parallel architectures\cite{flynn2005very}, and its practical \xwz{implementation}{realization} has evolved \xwz{alongside advancements}{along with advances} in processor design, particularly with the proliferation of SIMD instruction sets such as Intel AVX\cite{IntelAVX2}, ARM SVE~\cite{stephens2017arm}, and RISC-V V Extension\cite{cui2023risc}.}{} Vectorization plays a pivotal role in performance-critical domains by dramatically improving instruction-level parallelism. Consequently, automatic vectorization has become an indispensable component of contemporary compiler toolchains. 

\wuk{Two primary vectorization strategies are employed by mainstream compiler toolchains}{Modern compilers generally employ two primary vectorization strategies}, namely loop vectorization and Superword Level Parallelism (SLP) vectorization\cite{larsen2000exploiting}. Loop vectorization, the more established approach, transforms entire loop iterations into vector operations by exploiting cross-iteration parallelism. However, this method requires an extra verification given that loop-carried dependencies are absent. In contrast, SLP vectorization operates at the basic block level, identifying and grouping isomorphic scalar operations within loop bodies or straight-line code into vector instructions~\cite{porpodas18lslp,porpodas18vwslp,porpodas19snslp}.
Both\wuk{}{vectorization} approaches face significant limitations when dealing with real-world complex codes. Loop vectorization inherently fails to handle loop-carried dependencies, situations where iteration i+1 depends on results from iteration i (e.g., recurrence relations like prefix sums). SLP vectorization, while less constrained by cross-iteration dependencies, struggles with non-isomorphic operation patterns. 

The emergence of large language models (LLMs) and their ability to capture complex patterns have spurred research into LLM-based code optimization techniques. LLM-Vectorizer\cite{10.1145/3696443.3708929} demonstrates notable performance by transforming source code into vectorized implementations through intrinsic instructions.
However, \wuk{three limitations are exhibited}{it exhibits three limitations}: (1) inadequate handling of loop-carried dependencies that hinder effective parallelization, (2) generation of platform-specific constructs that compromise portability across architectures, and (3) production of low-readability code lacking maintainability.
In parallel, Self-Refine\cite{madaan2023self} employs iterative LLM-based refinement to progressively enhance code quality.
Nevertheless, this approach suffers from insufficient integration with compiler semantics, resulting in suboptimal optimizations due to the absence of compiler-guided metrics and target-specific optimization objectives. Both paradigms reveal fundamental challenges in integrating LLMs with compiler optimization frameworks. AlphaEvolve\cite{alphaevolve} also adopts an iterative approach to generate answers while possessing verification capabilities. Its main difference from Self-Refine lies in not dividing the iteration process into distinct feedback and refine phases. Additionally, AlphaEvolve similarly lacks the utilization of tools such as compilers.

\begin{figure*}[htbp]
    \centering
    \includegraphics[width=\textwidth]{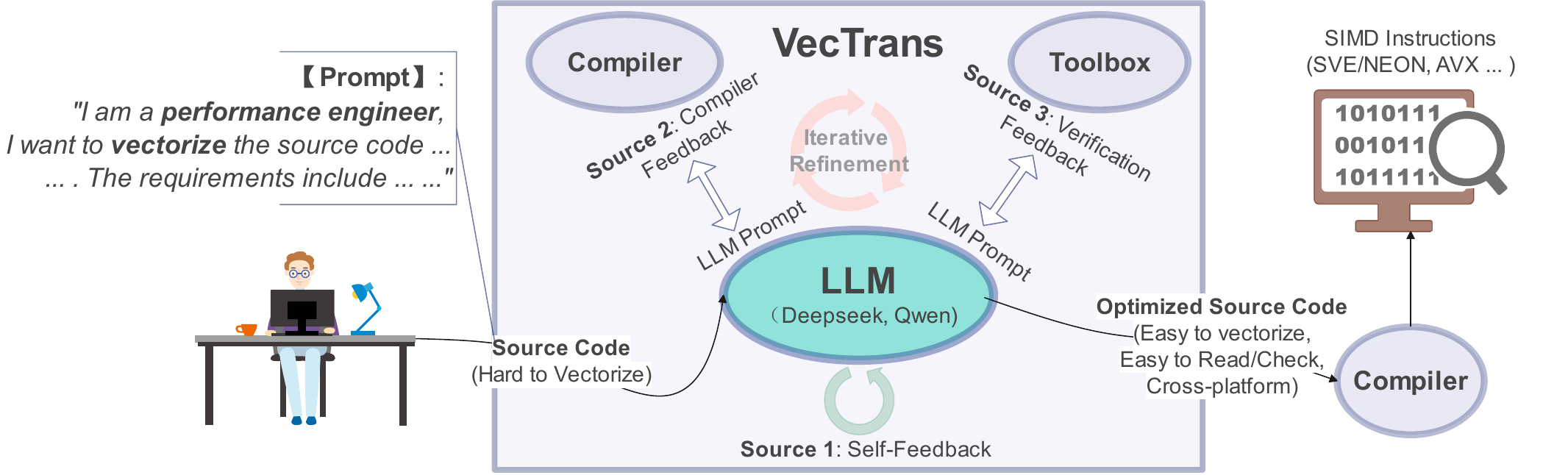}
    \caption{\wuk{The workflow of \NAME: \ding{182} \textbf{Dual-Phase Iteration} - Alternating feedback and refinement; \ding{183} \textbf{Tri-Source Diagnostics} - LLM self-evaluation, compiler-derived static analysis, and runtime validation; \ding{184} \textbf{Guided Code Transformation} - Employing transformations in the refinement phase informed by diagnostics to improve vectorization compatibility; \ding{185} \textbf{Adaptive Termination} - Dynamic termination when reaching either successful vectorization or optimization thresholds.}{} }
    \label{fig:overview}
\end{figure*}

\subsection{\wuk{Opportunities of LLM-assisted code transformation}{}}

\xwz{While compilers have made significant progress in auto-vectorization, they still face fundamental limitations in recognizing and transforming certain code patterns.}{} \wuk{As shown in Table \ref{tab:classification}, 50 out of 149 functions in the TSVC benchmark}{In the TSVC benchmark, 50 out of 149 functions} resist compiler auto-vectorization. \cl{Even for recently pioneering work related to LLM-Vectorizer\cite{10.1145/3696443.3708929} \wuk{--}{}which tries to integrate LLM, compiler, and verification into an \wuk{unified}{} AI agent framework, it also \wuk{miss substantial}{loses many} optimization opportunities,\wuk{}{possibly} due to the non-optimal interaction manner among different agent components.}{} \cl{}{To address the shortcomings\xwz{}{of the compiler}, LLM Vectorizer\cite{10.1145/3696443.3708929} has \xwz{pioneered the use of large language models}{done some pioneering work} to \xwz{}{greatly} improve the vectorization performance\xwz{, yet plenty of cases remain unaddressed}{. However, some cases cannot be covered}. }

\wuk{To illuminate the missed opportunity of auto-vectorization by code transformation, we present an example in Figure \ref{code:s1113}}{\cl{To better illuminate the motivation of our work, a representative example is}{} \xwz{illustrated}{demonstrated} in Figure \ref{code:s1113}}. The \xwz{original}{} source code as given in Figure \ref{code:s1113}(a) cannot be vectorized with traditional compilers even on \texttt{-O3} mode for two reasons\xwz{: (1)}{, first,} when \texttt{i=LEN\_1D/2}, \texttt{a[LEN\_1D/2]} is updated, which results in a read-after-write \xwz{(RAW)}{} dependency in subsequent iterations, and \xwz{(2)}{second,} \xwz{simultaneous}{} reading and writing an array {\it a} \xwz{within the loop}{at the same time} \xwz{makes vectorization unsafe}{is an insecure vectorization operation}, \xwz{causing the compiler to reject the transformation}{so the compiler rejects vectorization of this function}. \xwz{Upon further}{Through} analysis, it is found that \texttt{a[LEN\_1D/2]} is \tyl{a}{an} variable and is updated only once\xwz{, making it loop-invariant when the loop is split into two separate sections}{. Therefore, the loop is divided into two parts. Then, \texttt{a[LEN\_1D/2]} is an invariant in both loops}.
\wuk{Our evaluation reveals that LLM-Vectorizer\cite{10.1145/3696443.3708929} fails to produce vectorizable code across multiple iterations. This stems from two key limitations: (1) generated SIMD intrinsics consistently fail verification, and (2) the system lacks effective correction mechanisms. These findings suggest that LLM-based vectorization may benefit more from source-level transformations than low-level code generation, as LLMs appear more adept at handling high-level code patterns derived from their training data.}{}


\wuk{We summarize three key challenges in LLM-based auto-vectorization: (1) Suboptimal Component Interaction: Naive integration of LLMs with compiler toolchains leads to verification failures (Missing compiler feedback roughly halves performance) and missed optimization opportunities due to incoherent feedback loops. (2) Semantic-Preservation Dilemma: LLMs’ high-level pattern matching often violates compiler-required semantics (e.g., aliasing rules), while strict verification overly constrains optimization potential. }{}(3) Large language models manifest erratic convergence behavior during generative refinement iterations (e.g., oscillating between excessive vectorization attempts and conservative rollbacks), whereas traditional compiler heuristics lack the adaptive mechanisms needed to orchestrate their co-evolution toward convergent optimizations.


\section{\xwz{Design}{System Overview}}

\wuk{In this section, we propose \NAME,}{VecTrans is} an iterative refinement LLM optimization framework that uses feedback-refined iteration structures\cite{madaan2023self}. The feedback-refined iteration is a cyclical optimization process comprising two phases: the feedback phase and the refine phase.

During the feedback phase, information relevant to the optimized function is gathered, referred to as feedback information. This feedback is utilized to assess the correctness of the optimized function and to identify the potential for vectorization. The feedback information is derived from three sources, as illustrated in Figure 2:

\begin{itemize}
    \item \textbf{Source 1: Self-Feedback}. The basic feedback information comes from the capabilities of the LLM itself.
    \item \textbf{Source 2: Compiler Feedback}. Feedback information from the compiler, including compilation information and vectorization information. This information is used to\wuk{}{help the LLM} determine whether the currently generated optimized code can be compiled and vectorized successfully.
    \item \textbf{Source 3: Verification Feedback}. The feedback information is used to check whether the generated optimized code can pass the unit tests and formal verification. This helps the large language model determine the semantic correctness of the optimized code.
    
\end{itemize}

Following the feedback phase, a structured prompting strategy is applied during the refine phase to synthesize multi-source feedback into actionable optimizations. Specifically, the three feedback streams are systematically integrated with a vectorization-focused prompt template to guide the LLM's code regeneration.

\subsection{\wuk{}{Source 1: }Self-Feedback}
The self-feedback mechanism harnesses the LLM's analytical capabilities to evaluate optimized code along four dimensions: lexical correctness, syntactic validity, semantic equivalence, and vectorization potential. Through structured prompting, the LLM conducts a staged analysis process: (1) Lexical and Syntactic Validation: detecting tokenization and grammar errors; (2) Semantic Equivalence Checking: simulating unit test outcomes to predict runtime behavior without actual execution; (3) Vectorization Feasibility Assessment: analyzing loop structures to determine whether full vectorization is achievable; and (4) Optimization Opportunity Diagnosis:  identifying targeted transformations for semantically correct code. This comprehensive evaluation yields actionable insights that directly inform iterative optimization decisions.

\begin{figure}[htbp]
    \centering
    \includegraphics[width=0.42\textwidth]{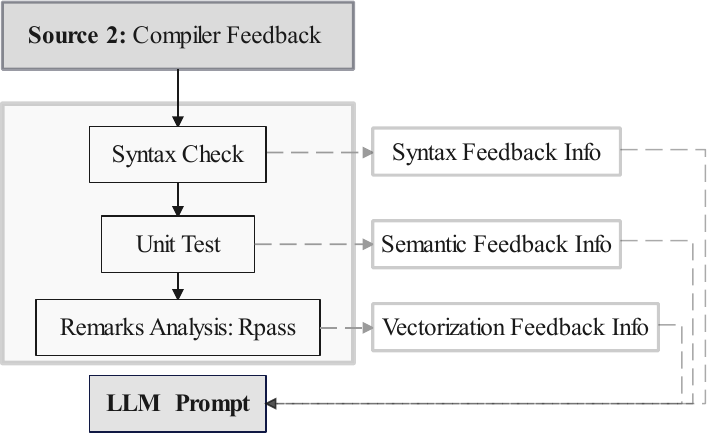}
    \caption{The compiler provides feedback through lexical/syntactic errors, semantic unit test results, and vectorization insights via Rpass flags. \wuk{}{This multi-source diagnostic evaluates code validity and vectorization potential.}}
    \label{fig:LLMFeedback}
\end{figure}

\subsection{\wuk{}{Source 2: }Compiler Feedback}


Compiler feedback information consists of three parts as shown in Figure \ref{fig:LLMFeedback}. In the first part, it conducts a compilation error check on the optimized code, capitalizing on the error-related information provided by the compiler. In the second part, it performs a semantic-equivalence analysis by comparing the source code and the optimized code through unit tests. In the third part, the Rpass and Rpass-analysis parameters of the compiler output vectorized information about the optimized code, including the positions of the loops that were successfully vectorized, the positions of the loops that were not successfully vectorized, and the reasons why the vectorization was not successful. These three types of information enable the LLM to assess whether the code is optimized and fully vectorized while preserving semantic correctness.


\begin{figure}[htbp]
    \centering
    \includegraphics[width=0.5\textwidth]{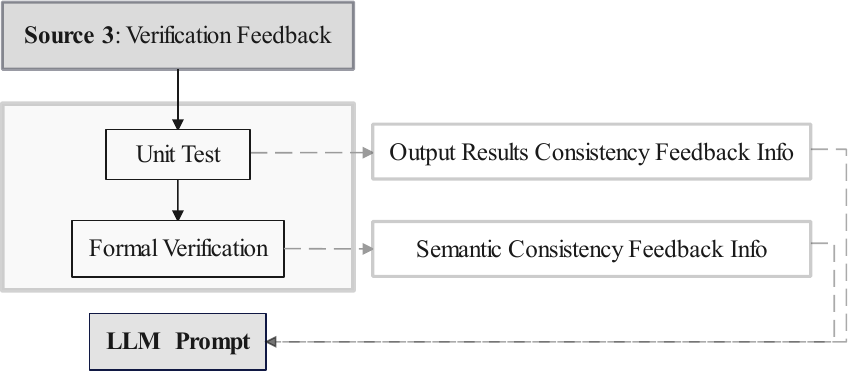}
    \caption{\wuk{Verification combines: \textbf{Unit Test} - Output-equivalence testing across representative inputs; \textbf{Formal verification} - Specification-based analysis of behavioral equivalence.}{}}
    \label{fig:CorrectVerify}
\end{figure}

\subsection{\wuk{}{Source 3: }Verification Feedback}
The verification feedback is divided into unit test information and formal verification information, shown in Figure \ref{fig:CorrectVerify}:

\textbf{Unit testing}: 
The unit testing phase verifies both syntactic correctness (via compilation) and semantic equivalence between optimized and original code. The framework leverages LLM-generated tests to validate semantic preservation through I/O comparisons. Tests are generated once during initialization and reused across iterations to minimize overhead. Test validity is ensured via two-step verification: (1) Baseline Equivalence: The optimized function is replaced with the original to confirm identical outputs, ensuring test reliability. (2) Test Sensitivity Check: For non-void returns, the original function is modified to return bitwise-negated results, replacing the optimized code to enforce a controlled mismatch. For void returns, the optimized code is replaced with an empty implementation to induce behavioral divergence. This approach reliably detects invalid test suites, with rare false positives only occurring for empty original functions—an edge case excluded due to negligible practical impact.

\textbf{Formal verification}: 
Following unit testing, the optimized code undergoes formal verification using Alive2\cite{10.1145/3453483.3454030}, where both original and optimized functions are compiled to IR and analyzed. Crucially, this verification occurs prior to vectorization, ensuring minimal divergence between IR versions and thus simplifying the verification process. Alive2 may produce four distinct outcomes: semantic equivalence (proceeding to performance testing), semantic mismatch, timeout, or verification error (triggering corrective feedback). Successful verification advances the pipeline, while failures initiate iterative refinement with diagnostic error information.



\subsection{Combination of \xwz{Feedbacks}{feedback information}}
In the refine phase, a comprehensive set of inputs is integrated, including the source code, the optimized code, feedback information, and relevant prior knowledge. This prior knowledge covers optimization techniques such as loop splitting, loop reordering, instruction reordering, and the use of temporary variables, along with simple illustrative patterns like instruction splitting (which requires independence between instructions) and iteration range splitting (applicable when loops exhibit phase-dependent behaviors).

This phase builds upon the feedback collected in the previous stage. The LLM takes the feedback data, source code, and current optimized code as input, and, guided by the embedded prior knowledge, iteratively refines the optimized code. The result is an improved code version, explicitly designed to enhance performance while preserving correctness and adhering to established coding conventions.

A notable characteristic of this phase is the strict constraint that the LLM generates only the optimized code, encapsulated with specific identifiers. This design choice simplifies downstream post-processing and facilitates seamless integration. Furthermore, by decoupling the feedback and refinement phases, the framework improves LLM efficiency and maintains a clear separation of concerns. This modular design enhances the scalability and automation potential of the optimization workflow by enforcing well-defined input-output interfaces for each phase.

\section{Evaluation}
\subsection{Experimental Methodology}
\textbf{System Configuration.} We conduct experiments on \wuk{}{two} Linux-based platforms with the following \wuk{two}{} specifications: 
\begin{itemize}
    \item \textbf{Arm Platform:} Kunpeng 920, 128-core@2.6GHz
    \item \textbf{X86 Platform:} Intel Xeon Gold 6132, 28-core@2.6GHz
\end{itemize}

\wuk{W}{For evaluation, w}e use GCC 14.2.0, Clang 20.0.0, BiSheng Compiler 3.2.0.1\cite{bisheng}, and Intel(R) oneAPI DPC++/C++ Compiler (ICX) 2023.2.0, with \wuk{compile}{optimization} flag \texttt{-O3 -ffast-math} and vectorization diagnostics enabled via \texttt{-Rpass\allowbreak=loop-\allowbreak vectorize\allowbreak \ -Rpass\allowbreak-analysis\allowbreak=loop-\allowbreak vectorize}. \wuk{W}{For our experiments, w}e utilize the Deepseek-V3 API and conduct comparative analyses against the Qwen2.5-Coder-32B-Instruct and Qwen2.5-72B-Instruct models.

\textbf{Benchmarks.} We evaluate a wide range of functions from TSVC\_2\cite{maleki2011evaluation} and real application function from Skia\cite{google2014skia}. The TSVC\_2 (Test Suite for Vectorization Compilers) benchmark suite systematically evaluates compiler capabilities in automatic vectorization through 149 test functions representing typical computational patterns. \wuk{Former}{Experimental} results reveal that 50 functions (approximately 33.6\%) resist effective vectorization by BiSheng Compiler\cite{bisheng} and mainstream compilers, primarily due to complex data dependencies, irregular memory access patterns, or unstructured control flows. The classification of these functions is shown in Table \ref{tab:classification}. These selected cases serve as critical evaluation targets, exposing limitations in current vectorization techniques while providing a standardized \wuk{test suite}{framework} for analyzing compiler optimization boundaries and guiding algorithmic improvements.

\textbf{Metrics.} \wuk{We assess framework effectiveness through three key metrics: (1) Coverage: Percentage of compiler-unvectorizable functions successfully transformed (primary success metric); (2) Speedup: Performance gain of vectorized functions (benefit metric); (3) Iteration Cost: Optimization rounds required (deployment cost metric).}{We consider the effectiveness of the framework from three dimensions: coverage, acceleration ratio, and iteration rounds. \tyl{All of the selected functions cannot be automatically vectorized by the compiler.}{The selected functions are all functions that cannot be automatically vectorized by the compiler.} Through the transformation of the framework, some functions can be automatically vectorized. The proportion of these functions, that is, coverage, is our primary consideration. The speedup of the test functions that can be successfully vectorized is \wuk{also}{another} \tyl{considered metric reflecting the benefits of the framework}{that we consider to reflect the benefits that the framework can bring}. Iteration rounds reflect the resources consumed by \tyl{conducting this framework for optimization}{using the framework for optimization}, including the number of iteration rounds needed for functions that \tyl{fail to be automatically vectorized}{cannot be automatically vectorized successfully}, and reflect the cost of deploying the framework in the compiler.}

\textbf{Parameter Setup.} The LLM selected for the experiment is Deepseek-V3\cite{liu2024deepseek}. The maximum number of iteration rounds is set to 20. Set the maximum number of tokens to 4096 and retain the default values for other parameters.

\begin{table}[htbp]
    \centering
    \caption{The classification of functions that cannot be vectorized in TSVC\_2.}
    \begin{tabularx}{.5\textwidth}{
    >{\raggedright\arraybackslash}X|
    >{\centering\arraybackslash}X}
    \toprule
       Category &  Function name\\
       \midrule
      unsafe dependent memory operations (13)  &s112, s1113, s114, s115, s116, s141, s212, s241, s242, s244, s1244, s281, s293\\
      \midrule
    \wuk{unable to}{could not} identify reduction variable (33)& s116, s123, s126, s141, s161, s211, s212, s1213, s221, s222, s231, s232, s233, s2233, s235, s2251, s256, s258, s261, s275, s277, s291, s292, s2111,  s318, s3110, s3112, s321, s322, s323, s341, s342, s343 \\
      \midrule
    \wuk{unable to}{cannot} identify array bounds (2)&s123, s1161\\
    \midrule
    \wuk{unable to}{could not} determine number of loop iterations (5)&s161, s277, s332, s481, s482\\
    \midrule
    instructions (e.g., call instruction) cannot be vectorized  (2) & s31111, s451\\
    \midrule
    loop contains switch statements (1)& s442\\
    \bottomrule
    \end{tabularx}
    \label{tab:classification}
\end{table}

\begin{figure*}
    \centering
    \includegraphics[width=\linewidth]{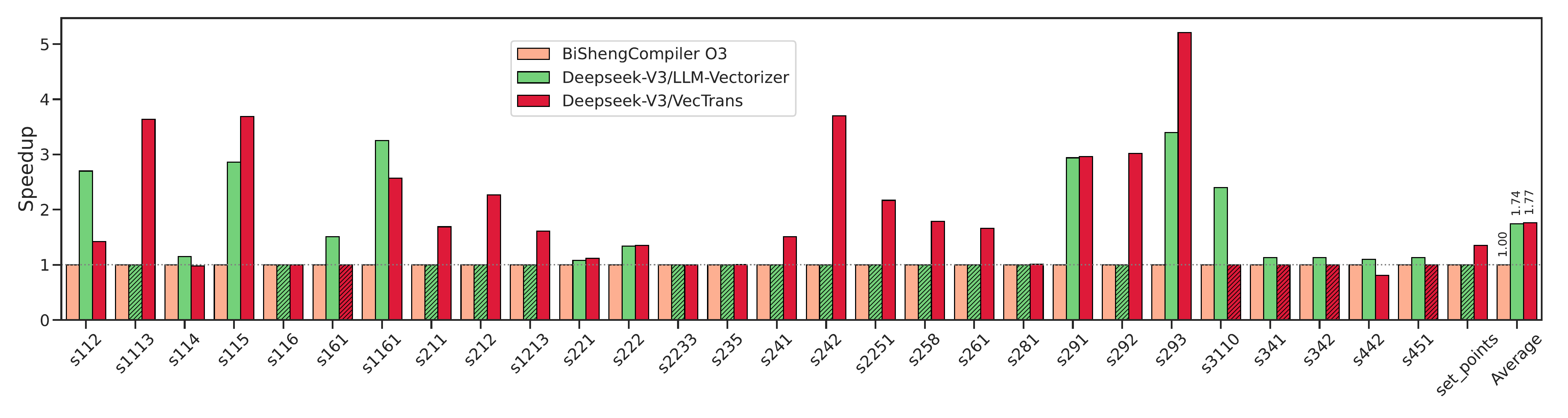}
    \caption{\NAME delivers a geometric mean speedup of 1.77×, marginally outperforming LLM-Vectorizer’s 1.74×. Shaded bars denote vectorization failures. The evaluation covers 29 functions, representing the union of vectorizable cases. Both frameworks use Deepseek-V3, with LLM-Vectorizer re-evaluated on Neon and Deepseek-V3 for consistency.}
    \label{fig:comparewithsota}
\end{figure*}

\subsection{\wuk{Speedup}{}}
    


\wuk{Our evaluation demonstrates that \NAME, when integrated with the BiSheng Compiler\cite{bisheng}, successfully vectorizes 24 functions (23 from TSVC and 1 from Skia) in our benchmark suite. As shown in Figure \ref{fig:comparewithsota}, \NAME delivers a 1.77x geometric mean speedup over the -O3 baseline, peaking at 5.21x for function s293.}{} When evaluated under ICX, our technique yields a 1.2× geometric mean speedup.

\wuk{However, results reveal two limitations: (1) two functions exhibit performance degradation (up to 19\% slowdown), and (2) four functions show negligible improvement (speedup \textless  5\%). A representative example is function s442, which achieves only a 0.81× speedup after transformation. This degradation stems from additional computations introduced during vectorization. Specifically, s442 contains a switch statement, and to enable automatic vectorization, the transformation replaces each branch with a condition-masked computation. For example, the transformed code computes every possible branch result and selects the correct one via condition multipliers (e.g., (condition) $\times$ (statement)). Microarchitectural analysis reveals three primary causes for these suboptimal cases:  (1) memory overhead from temporary array allocations, (2) increased cache miss rates due to non-contiguous memory access patterns, and (3) pipeline stalls caused by vector instruction scheduling conflicts. These findings highlight the critical trade-offs between vectorization potential and architectural constraints that must be carefully balanced in automatic vectorization frameworks.}{}

\subsection{Coverage}
\label{sec:coverage}

\wuk{Our framework demonstrates significant improvements in vectorization coverage, successfully vectorizing 23 of 50 TSCV functions (46\%), as well as complex real-world cases from Skia. This substantially outperforms LLM-Vectorizer's 28.8\% coverage on Neon ISA targets, highlighting the impact of limited architecture-specific training data in existing approaches.}{Our model can successfully vectorize 23 of the 50 TSCV functions, as well as a complex function from skia, coverage reached 46.2\%. The coverage of LLM-Vectorizer on Neon instruction sets is only 28.8\%. Limited Neon-specific training data in LLMs adversely impacts LLM-Vectorizer's operational effectiveness.}

Figure \ref{fig:comparewithsota} and Table \ref{tab:coverage} present the vectorization success rates across the categories in Table \ref{tab:classification}. Our framework achieves the highest efficacy (10/13 success) on functions previously unvectorizable due to unsafe memory dependencies. Reduction variable identification posed greater challenges, resulting in only 13 out of 33 functions being vectorized. The remaining four categories posed significant optimization barriers—only s1161 was successfully vectorized, with no additional functions from these groups being optimized. Figure \ref{fig:comparewithsota} also presents speedups for 29 functions, corresponding to the union of those vectorized by \NAME and LLM-Vectorizer. Notably, LLM-Vectorizer excels in handling conditional branches (right side of Figure \ref{fig:comparewithsota}), particularly in functions s3110 to s342, which compute array maxima or minima with irreducible control flow. These constructs typically resist conventional auto-vectorization and persist even after \NAME transformations. LLM-Vectorizer addresses this by merging multiple conditional checks into single vector instructions, though parallelism remains constrained, limiting achievable speedups.
Conversely, \NAME demonstrates superior performance on loops with cross-iteration dependencies (e.g., s2233–s281), where it systematically eliminates reducible dependencies through targeted transformations, enabling effective compiler auto-vectorization. While LLM-Vectorizer can occasionally discover dependency-breaking patterns during exploration, its lack of explicit dependency elimination strategies results in less consistent outcomes compared to the principled, dependency-aware approach of \NAME. Together, the two methods complement each other, collectively broadening the scope of vectorizable code.



Table \ref{tab:optimization_methods} shows the methods in which the LLM transformation function enables vectorization. Through these transformations, the compiler can more easily identify whether there is a dependency between loops, so that the compiler can vectorize a function that could not be vectorized originally. Among optimization techniques, loop splitting emerges as the most effective solution for eliminating false dependencies in loop structures. Instruction reordering and temporary variable insertion are employed to enhance compiler dependency analysis, which not only improves false dependency identification but also prevents unsafe memory accesses such as simultaneous array read-write operations. While branch elimination converts conditional jumps into arithmetic expressions, this approach inevitably introduces substantial redundant computations, thus generally offering no performance benefits.

\begin{table}[htbp]
    \centering
    \caption{Code Optimization Techniques and Descriptions}
    \label{tab:optimization_methods}
    \begin{tabularx}{.51\textwidth}{
    >{\raggedright\arraybackslash\hsize=0.55\hsize}X|
    >{\centering\arraybackslash\hsize=1.45\hsize}X}
        \toprule
        \textbf{Optimization} & \textbf{Description} \\
        \midrule
        Loop Splitting & 
        Decompose a single loop into multiple independent sub-loops to enable parallel execution. Commonly used for reducing loop-carried dependencies. \\
        
        \midrule
        Instruction Reordering/Loop Reordering & 
        Reorganize instruction execution sequence to eliminate false dependencies, or rearrange loop nesting order to enhance data locality. \\
        
        \midrule
        Temporal Variable & 
        Introduce short-lived variables to store recurrent computation results, reducing dependency caused by array access.  \\
        
        \midrule
        Branch Elimination & 
        Remove conditional statements like switch or if. \\
        \bottomrule
    \end{tabularx}
\end{table}

Beyond the successfully vectorized cases, the framework fails to transform 27 \wuk{}{additional} test functions. Most of these functions lack vectorization potential, and forced vectorization would lead to performance degradation. The failures occur due to:


\begin{itemize}
    \item \textbf{Iteration Limit Exceeded (9 cases).} The iteration upper bound is set to 20. If the LLM fails to vectorize the function after 20 iterations, the process terminates, and the original code is preserved.
    \item \textbf{Premature Optimization Assumption (2 cases). }In some cases, the LLM disregards compiler-reported vectorization constraints, incorrectly concluding that the code is already optimized. This leads to early termination without meaningful vectorization.
    \item \textbf{Loops with No Vectorization Benefit (15 cases).} The LLM identifies loops where forced vectorization offers no performance gain (e.g., due to loop-carried dependencies or conditional branches). Attempting vectorization would introduce redundant memory copies, degrading performance (deoptimization). Thus, the original code is preserved.
    \item \textbf{Semantic Errors (1 case: s481).}  A single failure (1/51 tests) produced semantically incorrect code despite passing unit tests and formal verification. This stems from insufficient test coverage and limitations in the verification toolchain. Such errors are critical and will be addressed in future work.
\end{itemize}

More specifically, the framework encounters vectorization challenges primarily due to four key patterns: (1) complex conditional branches (affecting 11 test cases), (2) unresolvable loop-carried dependencies (8 cases), (3) irregular memory access patterns from multi-dimensional arrays (6 cases), and (4) external function calls requiring inlining (2 cases). Among these, conditional branches constitute the dominant limiting factor, while array-induced memory access complexities and loop dependencies present secondary obstacles. The remaining cases involve function calls that necessitate source-level inlining—a transformation currently beyond the framework's scope.

\begin{table}[htbp]
    \centering
    \caption{The coverage and rounds under different configurations. The base model means that the framework is without formal verification, compiler feedback, and unit tests.}
    \begin{tabularx}{.5\textwidth}{
    >{\raggedright\arraybackslash\hsize=0.7\hsize}X|
    >{\centering\arraybackslash\hsize=0.3\hsize}X|
    >{\centering\arraybackslash\hsize=0.3\hsize}X|
    >{\centering\arraybackslash\hsize=0.3\hsize}X}
    \toprule
        Configuration &  Coverage &Geomean speedup&  Iteration rounds\\
         \midrule
         DeepSeek-V3 / VecTrans  & $\mathbf{46.2\%}$&1.77x & 8.760\\
         \midrule
         DeepSeek-V3 / LLM-Vectorizer\cite{10.1145/3696443.3708929} &28.8\% &1.74x& 13.39\\
         \midrule
         DeepSeek-V3 / base model& 17.3\%&1.57x &5.412\\
         \midrule
         Qwen2.5-32B / VecTrans & 34.6\% &1.96x& 13.940\\
         \midrule
         Qwen2.5-72B / VecTrans& 38.5\% & 1.87x&12.260\\
         \midrule
         DeepSeek-V3 \wuk{w/o}{/Without} formal verification& 32.7\%&1.76x&8.489\\
         \midrule
         DeepSeek-V3 \wuk{w/o}{/Without} compiler feedback&21.2\% &1.74x&5.213\\
         \midrule
         DeepSeek-V3 \wuk{w/o}{/Without} unit test&25.0\% &1.52x&9.656\\
         \bottomrule
    \end{tabularx}
    
    \label{tab:coverage}
\end{table}

\begin{figure*}[htbp]
    \centering
    \includegraphics[width=\linewidth]{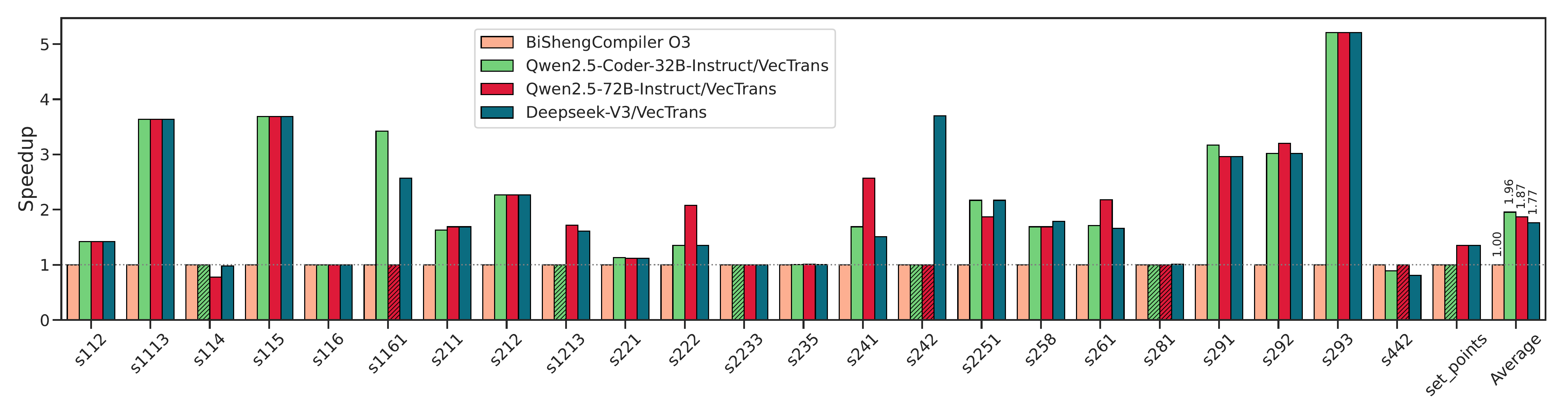}
    \caption{The graph shows the speedup of functions that can be successfully vectorized when the framework uses different LLMs. Columns with slashes indicate that vectorization cannot be successful. Three LLMs, Qwen2.5-Coder-32B-Instruct\cite{hui2024qwen2}, Qwen2.5-72B-Instruct\cite{yang2024qwen2} and Deepseek-V3, are used here. The speedup is of the final generated optimized function compared to the original function using BishengCompiler at \texttt{-O3 -ffast-math} flag.}
    \label{fig:diffmodel}
\end{figure*}

\subsection{\wuk{Iteration Rounds}{}}


The framework requires 8.76 iterations per function on average, with 9 functions reaching the 20-iteration ceiling. Successful vectorization demonstrates significantly greater optimization efficiency at 4.96 iterations (43.4\% reduction from baseline).


Token consumption analysis reveals 2.5k input/600 output tokens per iteration. Given the average 21.9k input/5.3k output tokens per function and DeepSeek-v3's pricing of \$0.27/M input and \$1.10/M output tokens, the vectorization cost per function averages \$0.012.

\subsection{Breakdown Analysis}

To understand the impact of the different components of the framework on VecTrans, we evaluated the impact of formal verification, compiler feedback, and unit test on the overall framework.


Table \ref{tab:coverage} presents quantitative evidence of component effectiveness through coverage metrics and iteration dynamics across architectural variants. Our ablation study reveals significant performance degradation when disabling individual components: removing compiler feedback causes the most substantial coverage drop (46.2\% → 21.2\%, -54.1\%), followed by unit test exclusion (46.2\% → 25.0\%, -45.9\%), with formal verification removal showing the least impact (46.2\% → 32.7\%, -29.2\%). The baseline configuration (without all three components) achieves only 17.3\% coverage, demonstrating their cumulative effectiveness through 2.67× improvement over the base model.

Compiler feedback enhances the framework in three key ways: (1) Helps LLM to determine whether a compilation error occurs. If a compilation error occurs, the optimized function needs to be regenerated. (2) Helps LLM determine whether vectorization is successful. If vectorization has not been performed, LLM needs to analyze whether vectorization opportunities exist and provide conversion suggestions. (3) Assists the LLM in identifying transformations that enable automatic vectorization by the compiler. The compiler gives the loop position that cannot be vectorized and the reason why it cannot be vectorized.

Therefore, when the compiler feedback is removed, the LLM needs to rely on its analysis capability to determine whether the current optimized function can be automatically vectorized. When the LLM mistakenly considers that the current function can be automatically vectorized, the LLM chooses to end the iteration and output code. In fact, the current function is not successfully vectorized. In experiments, this is common, so when the compiler feedback is removed, the coverage decreases, and the number of iteration rounds decreases, because the LLM considers that the current optimized function has been fully optimized in the first few iteration rounds and selects to output the code.

Unit test and formal validation are similar in that they are used to detect the semantic correctness of the generated code, where unit test is more important than formal verification in that it can intercept more incorrect code because formal verification will fail in some complex situations. The unit test is more applicable, but it cannot cover all cases due to the limited input range. Therefore, formal verification is needed as a supplement.

The removal of different components has an impact on the coverage rate and iteration rounds, but has little impact on the acceleration ratio because the code optimization methods used by the same LLM are similar, and the generated code is basically the same. Except that a few functions may reduce the acceleration ratio due to redundant operations, the acceleration ratio remains the same in other cases.



\begin{figure}[htbp]
    \centering
    \includegraphics[width=\linewidth]{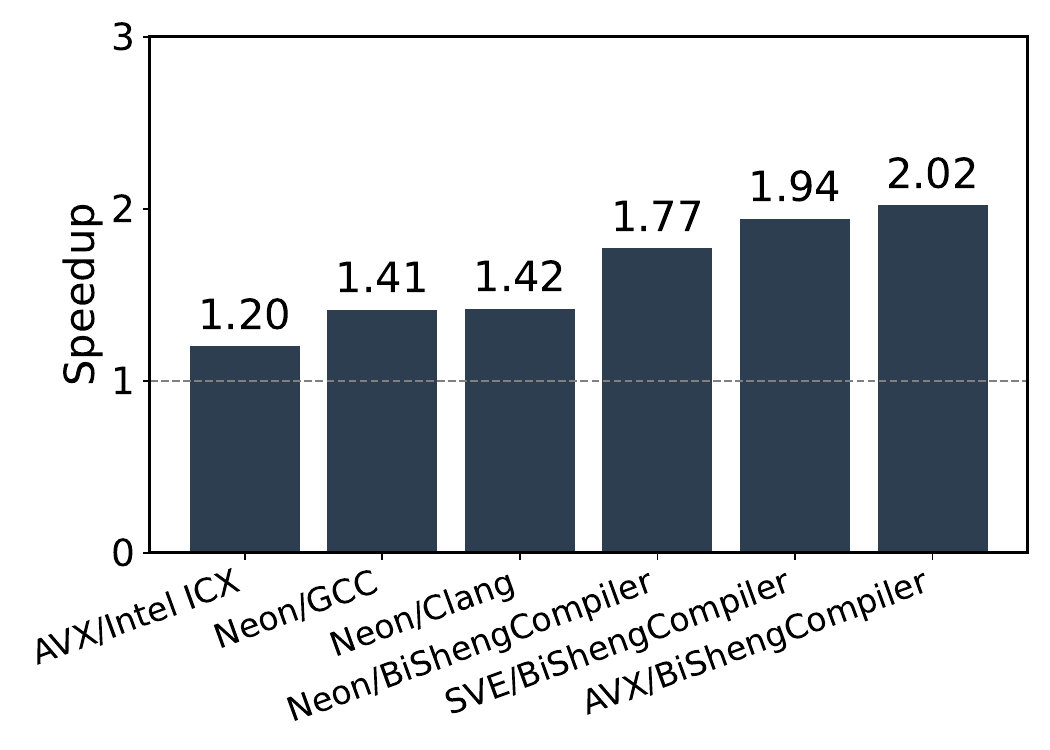}
    \caption{Geometric mean speedup of functions optimized by VecTrans across different configurations. Each bar reports the speedup relative to the \texttt{O3} baseline of the corresponding compiler and hardware backend.}
    \label{fig:sensivity}
\end{figure}

\subsection{Sensitivity Studies}
We verify the robustness of the framework from three dimensions: model, compiler, and architecture. Only one parameter is changed at a time, and the rest remain unchanged. We have verified that the framework works across different LLMs, compilers, and architectures.

\textbf{Sensitivity of different models.} We \tyl{evaluate}{evaluated} the usability of the framework on three LLMs, Qwen2.5-Coder-32B-Instruct, Qwen2.5-72B-Instruct, and Deepseek-V3. In addition, small-scale tests are performed on smaller models. It is found that the instruction compliance capability of smaller models is poor and the normal iteration process cannot be completed. The Deepseek-R1 model is also tested on a small scale, which can produce correct results in fewer iterations, but takes too long.

Table \ref{tab:coverage} presents the coverage rate and iteration rounds for the three base models. As model parameters decrease, we observe a reduction in coverage (DeepSeek-V3: 46.2\%, Qwen2.5-72B: 38.5\%, Qwen2.5-32B: 34.6\%) and an increase in iteration rounds (8.76, 12.26, and 13.94, respectively). This trend suggests that smaller models exhibit weaker capability, failing to transform some test functions successfully. Consequently, they require more iterations, with some cases failing to produce correct results within the 20-round limit.

Although the capabilities of the framework are affected by the base model, the coverage of small models is higher than that of large models without framework components. Therefore, the framework is robust on different models and can be migrated to different LLM.

Figure \ref{fig:diffmodel} shows the speedup of different model generation functions. The speedup of Qwen2.5-Coder-32B-Instruct is 1.96 times, and that of Qwen2.5-72B-Instinct is 1.87 times. The speedup of Deepseek-V3 is 1.77 times. Even for different LLMs, the generated functions are similar, and the speedup is similar. Therefore, this framework may help identify potential compilation optimization points, that is, summarize the characteristics of the generated functions. The optimization direction of automatically vectorized can be found.

\textbf{Sensitivity of different compilers.}
Here we explore the impact of different compilers on the optimized functions, and we test the speedup of GCC, Clang, ICX, and BiSheng Compiler\cite{bisheng}, respectively. The GCC and Clang versions used here are the latest versions, which represent the best compilers available today. ICX is Intel-optimized for superior CPU/GPU performance on Intel hardware, outperforming GCC/Clang in Intel-centric environments.  BiSheng Compiler is a compiler optimized for Kunpeng series chips.
GCC, Clang, and BishengCompiler are tested on Kunpeng 920, while ICX is tested on Intel CPU. The Kunpeng 920 uses the Neon instruction set, and Intel CPU employs the AVX instruction set.
The speedup of functions compiled using the flag \texttt{-O3} \texttt{-ffast-math} on different compilers is shown in Figure \ref{fig:sensivity}.

The geomean speedup of the optimized function is 1.42 times on Clang, 1.41 times on GCC, and 1.77 times on BiSheng Compiler\cite{bisheng}. The code generated for the s112 function is significantly worse on GCC and Clang than it is on BiSheng Compiler\cite{bisheng} because the latest versions of GCC and Clang can vectorize the original s112 function. In s1161 and s242, the speedup of GCC is lower than that of the other two compilers. In s293, the speedup of Clang is lower than that of the other two compilers. This indicates that different compilers optimize functions differently. However, in most cases, the optimization of these functions is effective for all three compilers, and can achieve good results. The speedup achieved with ICX was 1.2×, which is lower than that of the other three compilers. This discrepancy arises not only from differences in test hardware (Kunpeng 920 utilizing Neon instructions vs. Intel CPUs leveraging AVX) but also because ICX incorporates architecture-specific optimizations for Intel Xeon CPU, resulting in higher vectorization coverage that reduces its geomean speedup gains in this cross-platform comparison.


\textbf{Sensitivity of different architecture.}
Different vector instruction sets also have a great impact on optimized function performance. In the x86 architecture, AVX instruction sets are usually used. In the ARM architecture, NEON instruction sets and SVE instruction sets are used. We will explore the effects of different vector instruction sets on the performance of generating optimized functions on different architectures.

As shown in Figure \ref{fig:sensivity}, the geomean speedup is 2.02 times on the AVX instruction set, 1.77 times on the Neon instruction set, 1.94 times on the SVE instruction set, and on the AVX instruction set, the speedup of the s1161 function can reach 16.8 times. It can be observed that, on different vector instruction sets, there is no obvious regularity in the speedup of the function, but the effect is better on the AVX instruction set and the SVE instruction set. The effect comes from the length of the vectorized register on the one hand, and the vectorized implementation mode on the other hand.

\begin{figure}[htbp]
\centering
\begin{subfigure}[t]{0.48\textwidth}

\begin{lstlisting}[   % 进行参数设置
 language=C, % 设置语言
 basicstyle=\ttfamily, % 设置字体族
 breaklines=true, % 自动换行
 keywordstyle=\bfseries\color{DodgerBlue3}, % 设置关键字为粗体，颜色为 DodgerBlue3,
 columns=flexible
]  
void s481(...) {
    for (int nl = 0; nl < iters; nl++){
        for (int i = 0; i < LEN_1D; i++){
            if (d[i] < (float)0.) return;
            a[i] += b[i] * c[i];
        }
    }
}

\end{lstlisting}
\caption{Original code that performs updates to \texttt{a} until a negative value is detected in \texttt{d}, exiting immediately afterwards.}
\end{subfigure}

\begin{subfigure}[t]{0.48\textwidth}
\begin{lstlisting}[   % 进行参数设置
 language=C, % 设置语言
 basicstyle=\ttfamily, % 设置字体族
 breaklines=true, % 自动换行
 keywordstyle=\bfseries\color{DodgerBlue3}, % 设置关键字为粗体，颜色为 DodgerBlue3,
 columns=flexible
]  
void s481_opt(...) {
    for (int nl = 0; nl < iters; nl++){
        int early_exit = 0;
        for (int i = 0; i< LEN_1D; i++)
            if (d[i] < (float)0.){
                early_exit = 1;
                break;
            }
        if (early_exit) return;
        for (int i = 0; i < LEN_1D; i++)
            a[i] += b[i] * c[i];
    }
}
\end{lstlisting}
\caption{Incorrect optimization that hoists the negative check, altering the original semantics by potentially skipping valid updates.}
\end{subfigure}

\caption{Failure case: The transformation incorrectly hoists the termination condition, potentially omitting valid updates that would occur before negative values in the original execution.}
\label{code:failureCase}
\end{figure}

\subsection{\wuk{Case Study}{}}

\subsubsection{\textbf{Failure Case Analysis}}

As shown in Figure \ref{code:failureCase}, test case s481 generated erroneous code that passed both unit testing and formal verification. This section analyzes how this test case circumvented correctness verification. First, the test case passed unit testing due to randomly generated inputs. The large language model's generated test functions predominantly contained positive values in their random inputs, whereas triggering the conditional branch in this function requires negative values in the d-array. Consequently, the unit test inputs failed to cover this scenario, resulting in the function erroneously passing verification. This limitation can be mitigated by constraining the model to generate inputs with improved coverage characteristics. Second, this test case passes Alive2's verification because Alive2 can only perform bounded loop checking. It operates by unrolling loops in the target function for a fixed number of iterations and verifying the semantics of the unrolled version. Consequently, Alive2's soundness is constrained by this bounded verification approach, meaning it cannot guarantee correctness under all possible scenarios.

\subsubsection{\textbf{Real World Application}}

\wuk{In this section, we use real-world function from Skia\cite{google2014skia}, as shown in Figure \ref{code:complex}, to show how \NAME works in detial}{The framework has the ability to handle complex functions from real-world applications(Skia), as shown in Figure \ref{code:complex}}. Skia is an open-source 2D graphics rendering engine developed by Google, extensively utilized in projects such as Chrome, Android, and Flutter. It delivers high-performance support for vector graphics, text rendering, image processing, and GPU-accelerated rendering. In the source code, the presence of loop-carried dependencies in the dst array prevents vectorization. In the optimized code, the temporary array is used to temporarily store intermediate results, and loop splitting is used to separate the parts with loop dependencies, so that the remaining loops without dependencies can be vectorized successfully. After optimization by \NAME, this function achieves a 1.35x speedup on the Neon instruction set when compiled with BiShengCompiler.

\begin{figure*}[ht] 
\centering
\begin{subfigure}{0.49\textwidth}
\begin{lstlisting}[   % 进行参数设置
 language=C, % 设置语言
 basicstyle=\ttfamily, % 设置字体族
 breaklines=true, % 自动换行
 keywordstyle=\bfseries\color{DodgerBlue3}, % 设置关键字为粗体，颜色为 DodgerBlue3,
 columns=flexibl,
]  
void set_points(...) {
    /* ... */
    for (int i = 0; i < divCount; i++){
        src[i + 1] = divs[i];
        int srcDelta = src[i+1] - src[i];
        float dstDelta;
        if (srcFixed <= dstLen){
            dstDelta = isScalable ? scale * srcDelta : srcDelta;
        } else {
            dstDelta = isScalable ? 0.0f : scale * srcDelta;
        }
        dst[i + 1] = dst[i] + dstDelta;
        isScalable = !isScalable;
    }
    /* ... */
}
\end{lstlisting}
\caption{Original implementation with data-dependent control flow} 
\end{subfigure}
\hfill
\begin{subfigure}{0.49\textwidth}
\begin{lstlisting}[   % 进行参数设置
 language=C, % 设置语言
 basicstyle=\ttfamily, % 设置字体族
 breaklines=true, % 自动换行
 keywordstyle=\bfseries\color{DodgerBlue3}, % 设置关键字为粗体，颜色为 DodgerBlue3,
 columns=flexibl
]  
void set_points_opt(...) {
    /* ... */
    for (int i = 0; i < divCount; i++){
        src[i + 1] = divs[i];
        srcDelta[i] = src[i+1] - src[i];
        if (srcFixed <= dstLen){
            dstDelta[i]=isScalable?scale*srcDelta[i]:srcDelta[i];
        } else {
            dstDelta[i] = isScalable ?0.0f : scale * srcDelta[i];
        }
        isScalableArray[i] = isScalable;
        isScalable = !isScalable;
    }
    for (int i = 0; i < divCount; i++)
        dst[i + 1] = dst[i] + dstDelta[i];
    /* ... */
}
\end{lstlisting}
\caption{Optimized version enabling vectorization.}
\end{subfigure}
\caption{Optimization example in Skia: Eliminating dependencies via loop splitting and temporary buffering.}
\label{code:complex}
\end{figure*}

\section{Related Works}

\wuk{}{This section provides an overview of related work on LLM for compiler.}

\textbf{LLM for compilers.} \wuk{}{In recent years, there has been a lot of work using LLMs to optimize compilers, focusing on different aspects of code work.}
Recent work demonstrates the growing role of LLMs in compiler-related tasks. Armengol-Estapé et al.\cite{armengol2024slade} demonstrate an innovative decompilation using a small language model to ensure correctness while improving code readability. Program repair is another study that has been highlighted. Bouzenia et al.\cite{bouzenia2024repairagent} present a method of using LLMs to automatically repair programs, using iterative interaction, and show how large models interact with tools. Iterative interaction framework\cite{bi2024iterative}\cite{madaan2023self} is widely used in LLM applications to improve the ability of generation. In addition to optimizing code directly with LLM, another idea is to generate transform pass that can optimize code. Cummins et al.\cite{cummins2024don} use the way of chain of thought to generate transform pass, thus improving the accuracy of LLM optimization. In addition, the LLM is used to predict the pass list, which is another way to avoid code correctness verification\cite{cummins2023large}. By adjusting the pass list, the performance of the program can be greatly changed.

\textbf{\wuk{ML-based automatic}{ML for} vectorization.} \wuk{}{Machine learning techniques are now being used to improve automatic vectorization.}Kevin et al.\cite{stock2012using} replace the cost model of automatic vectorization with a machine learning model to help the compiler make vectorization decisions. Ameer et al.\cite{haj2020neurovectorizer} deploy a reinforcement learning method in the compiler to predict the optimal vectorization factor, which improves the performance of loop vectorization. Taneja et al.\cite{10.1145/3696443.3708929} propose an innovative approach to vectorization, using LLM to generate source code that uses Intrinsic instructions, thus completing vectorization at the source level and verifying the semantic correctness of the vectorized code.

\textbf{Classical vectorization \wuk{method}{}.}\wuk{}{The research on automatic vectorization has been a hot topic.} Franchetti and Püschel\cite{franchetti2008generating} use the term rewriting technique to vectorize small matrix kernels. Thomas et al.\cite{thomas2024automatic} use equality saturation to automatically generate high-quality vectorizing compiler for digital signal processors, creating a new idea for automatic vectorization.

\section{Summary}


We present a source code transformation framework leveraging large language models to restructure code into forms more amenable to automatic vectorization. The framework maximizes the compiler’s vectorization capabilities while employing unit tests and formal verification to ensure the correctness of the transformed code. By integrating LLMs with traditional compilers, the framework constrains the role of LLMs to inference-driven transformations, reducing the risk of semantic errors. This design preserves the reliability of conventional compiler vectorization while enhancing its applicability through learned code restructuring. We believe this approach opens new directions for integrating LLMs into compiler optimization pipelines. In future work, we plan to extend this framework to support additional optimizations and further strengthen its correctness guarantees.


\end{document}

%% file: preamble.tex
\usepackage[switch]{lineno}
\usepackage{xspace}
\usepackage{tabularx}
\usepackage{listings}
\usepackage[dvipsnames,x11names]{xcolor}
\usepackage{subcaption}
\usepackage{float}
\usepackage{stfloats}
\usepackage{cite}
\usepackage{amsmath,amssymb,amsfonts}
\usepackage{algorithmic}
\usepackage{graphicx}
\usepackage{textcomp}
\usepackage{balance}
\usepackage{booktabs}
\usepackage{pifont}
\usepackage[colorlinks,
linkcolor=black,
anchorcolor=black,
citecolor=black,
urlcolor=black
]{hyperref}
\usepackage{pbalance}